\documentclass[12pt,a4paper,twoside]{article}
\usepackage{epsfig}
\def\baselinestretch{1.1}
\def\arraystretch{0.9}
\newlength{\figwidth}
\newlength{\figheight}
\def\gaga{$\gamma \gamma$}

\def\epem{$e^+ e^-$}

\title{\vspace{1.5cm} CompAZ: parametrization of the luminosity spectra 
      for the photon collider.}

\author{Aleksander Filip \.Zarnecki \\
{\small\it Institute of Experimental Physics, Warsaw University,} \\
{\small\it Ho\.za 69, 00-681 Warszawa, Poland} \\
{\small\it E-mail: zarnecki@fuw.edu.pl}
       }

\begin{document} 

\maketitle 

\begin{abstract}

A simple model, based on the analytical formula for the Compton scattering,
is proposed to describe the realistic photon-energy spectra 
for the Photon Collider at TESLA. 
Parameters of the model are obtained from
the full simulation of the beam by V.Telnov, 
which includes nonlinear corrections and
contributions of higher order processes. Photon energy distribution and 
polarization, in the high energy part of the spectra, are well reproduced. 
Our model can be used for a Monte Carlo simulation of gamma-gamma events at 
various energies and for direct cross-section calculations.

\end{abstract}

\thispagestyle{empty}

%
%

\section{Introduction}
\label{sec:intro}

Photon Collider has been  proposed as a natural
extension of the \epem\ linear collider project TESLA \cite{tdr_pc}.
High-energy photons   
can be obtained using Compton
backscattering of the laser light off the high-energy 
electrons \cite{GKST81,GKST83,GKST84,TEL90,TEL95}.  
The physics potential of the Photon Collider is very
rich and complementary to the physics program of \epem\
and hadron colliders \cite{tdr_pc}.
It is the ideal place to study 
the mechanism of the electroweak symmetry breaking (EWSB) and
properties of the Higgs boson.
Precision measurements at the Photon Collider may open ``new windows''
to the physics beyond the Standard Model.
However, the precise measurements are only possible if the energy spectrum
of colliding photons is well understood.
A detailed simulation of the \gaga\  luminosity spectra at the Photon Collider
at TESLA has become available recently \cite{TEL95,TEL01}.
In this paper a simple model based on these results is proposed.
%


\section{Luminosity spectra}
\label{sec:simul}

If the laser beam density is small and
the primary electron beams are sufficiently wide, so that effects
related to the photon scattering angle can be neglected,
the energy spectrum of the photon beams colliding in the Photon Collider
could be calculated directly from the Compton scattering cross section
(see appendix \ref{sec:compton}).
However, these assumptions will not be fulfilled at the Photon Collider
as we need both very powerful lasers and strongly focused electron beams
to get high luminosity.
To find the energy spectra of colliding photons, 
with realistic assumptions about the laser system and the electron beams,
detailed simulation programs has been prepared \cite{TEL95,YOK}.

New samples, with high statistics of the simulated \gaga\ events, have
been generated recently by V.Telnov \cite{TEL01}.
They are based on the beam and laser system layout and parameters, 
as proposed for the Photon Collider at TESLA \cite{tdr_pc}.
Electron beam energies of 100, 250 and 400 GeV have been considered.
Basic parameters used in the simulation
are listed in Table \ref{tab:telnov}.
More details of the simulation can be found in \cite{tdr_pc,TEL01,TEL00}.
\begin{table}[bp]
{
\renewcommand{\arraystretch}{1.3}
{
\begin{center}
\begin{tabular}{l c c c} \hline
$E_e$ [GeV] & 100 & 250 & 400   \\ \hline
$\lambda_L$ [$\mu m$] & 1.06 & 1.06 & 1.06 \\
$E_0$ [eV] & 1.17 & 1.17 & 1.17 \\
$x$  & $1.8$ & $4.5$ & $ 7.2$ \\ \hline  
$\sigma_{x}$ [nm] & 140 & 88 & 69  \\  
$\sigma_{y}$ [nm] & 6.8 & 4.3 & 3.4  \\ 
$\sigma_{z}$ [mm]& 0.3 & 0.3 & 0.3  \\  \hline   
$f_{rep}$ [kHz]& 14.1 & 14.1 & 14.1  \\
$\gamma \epsilon_{x/y}/10^{-6}$ [m$\cdot$rad] & 2.5/0.03 & 2.5/0.03 & 
2.5/0.03 \\
$\beta_{x/y}$ [mm] at IP& 1.5/0.3 & 1.5/0.3 & 1.5/0.3 \\
b [mm] & 2.6 & 2.1 & 2.7 \\
$L_{e e}\, (geom)$ [$10^{34}\,cm^{-2} s^{-1}$] & 4.8 & 12 &  19 \\  \hline
\end{tabular}
\end{center}
}
\caption{Parameters of  the Photon Collider based on TESLA. 
Listed for different electron beam energies $E_e$ are: 
laser wave length $\lambda_L$,  laser photon energy $E_0$ 
and resulting $x$ parameter values (see section \ref{sec:model}
and appendix \ref{sec:compton});
horizontal, vertical and longitudinal electron bunch sizes 
$\sigma_{x}$, $\sigma_{y}$ and $\sigma_{z}$; 
average repetition rate, 
normalized beam emittances $\gamma \epsilon_{x/y}$,
$\beta$-functions $\beta_{x/y}$ , 
distance between the conversion and interaction point and
estimated $e^- e^-$ luminosity. }
\label{tab:telnov}

}
\end{table}   

\begin{figure}[tbp]
  \begin{center}
  \epsfig{figure=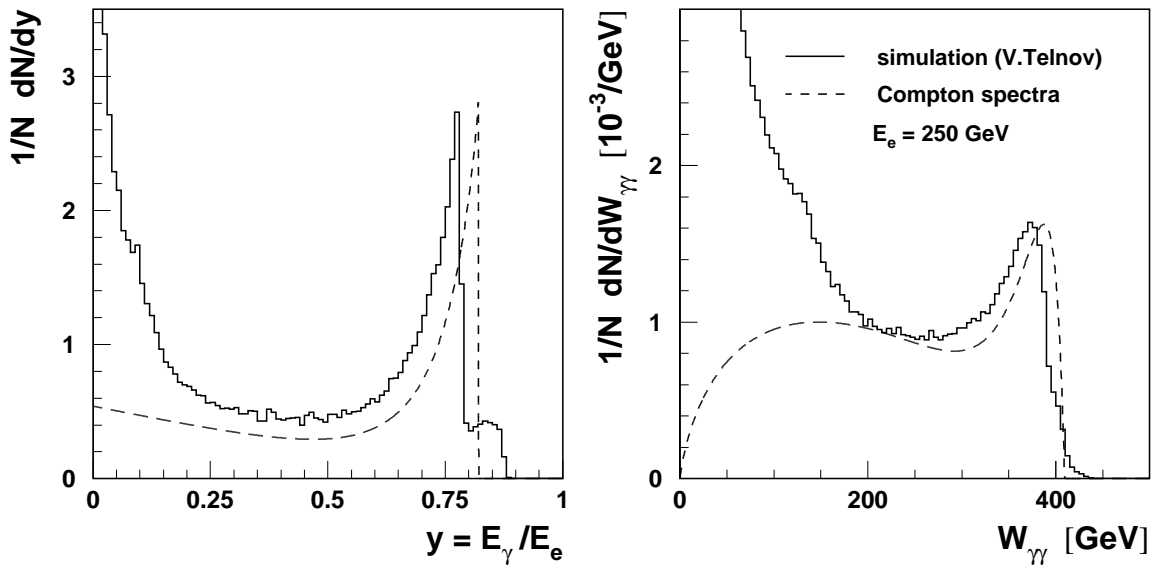,width=\figwidth,clip=}
  \end{center}
  \caption{Energy distribution for photons (left plot)
           and the $\gamma \gamma$ center-of-mass energy 
           distribution (right plot) from full simulation of 
           luminosity spectrum by V.Telnov \cite{TEL01} (solid line),
           compared to expectations for the simple Compton scattering
           (dashed line).
           For better comparison of shape, 
           Compton spectra is scaled to the same height
           of the high energy peak.
           }
  \label{fig:spectra}
\end{figure}

Shown in Fig.~\ref{fig:spectra} are the distribution of the
colliding photon energy ratio to the primary electron beam energy, 
$ y = E_\gamma / E_e $, and the distribution of 
the \gaga\ center-of-mass energy distribution, $W_{\gamma \gamma}$.
Results obtained from the simulation of luminosity spectrum \cite{TEL01},
for electron beam energy of 250 GeV, are compared with
the distributions expected from the simple Compton scattering
(lowest order QED), as given by formula (\ref{eq:compt}) 
(see appendix \ref{sec:compton}).
Realistic simulation indicates that a large fraction of colliding
photons will have small energies.
Also the maximum of the high-energy peak is shifted towards lower
energies.

\begin{figure}[tbp]
  \begin{center}
 \epsfig{figure=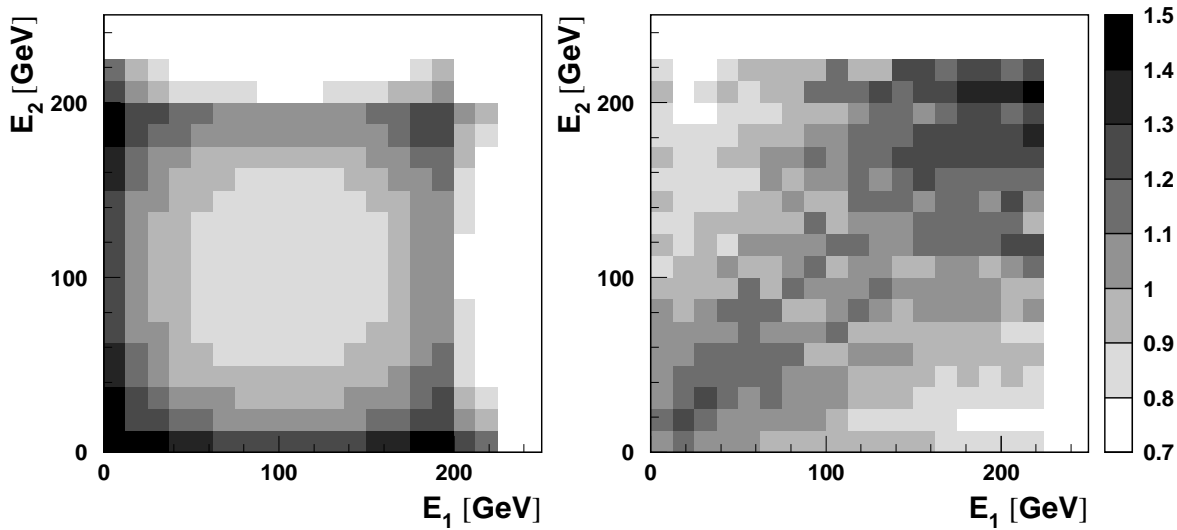,width=\figwidth,clip=}
  \end{center}
  \caption{Two-dimensional energy distribution for two colliding photons, 
           obtained from the full simulation by V.Telnov \cite{TEL01} 
          (left plot) and the ratio of this distribution to 
           the simple product of two one-dimensional
           energy spectra (right plot).
           }
  \label{fig:spectra2}
\end{figure}

Shown in Fig.~\ref{fig:spectra2} (left plot) is
the two-dimensional energy distribution for the colliding photons,
as obtained from the simulation \cite{TEL01}, 
for electron beam energy of 250 GeV.
Also included in Fig.~\ref{fig:spectra2} (right plot) is
the energy correlation between two photons calculated as
the ratio of the two-dimensional energy distribution
to the product of two one-dimensional energy spectra.
Energies of colliding photons are clearly correlated. 
Majority of collisions involve photons with similar energies
(large values of the ratio along the diagonal).
Collisions involving one low-energy and one high-energy photon
are suppressed (the ratio less than 1 in the left-upper 
and right-lower corner of the plot).
This demonstrates that the correlation between the angle of 
Compton backscattering and the photon energy is important 
and has to be taken into account.

\section{Model}
\label{sec:model}

Dedicated studies have been performed 
in order to understand the differences 
between the photon energy spectra obtained from the collider simulation 
by V.Telnov  \cite{TEL01}
and the spectra expected for the simple Compton scattering (\ref{eq:compt}).
To describe simulation results
following effects are taken into account:
nonlinear effects due to high density of the laser beam, 
correlation between photon energy and
the scattering angle, electron rescattering and
scattering involving two initial state laser photons.

\subsection{Nonlinear effects}

For very high density of the laser beam nonlinear QED effects
become important.
The field of the electromagnetic wave can significantly influence
the motion of an electron.
The effect can be described as an effective increase in the electron mass
$m_e^2 \rightarrow m_e^2 ( 1 + \xi^2)$, where $\xi^2$ is the parameter
describing the nonlinear effects, proportional to the photon density 
in the laser beam \cite{Galynskii}.  
For the Compton scattering nonlinear effects result in
an effective rescaling of the parameter $x$, describing the
photon energy spectra:  
\begin{eqnarray}
x = \frac{4 E_0 E_e}{m_e^2} & \longrightarrow & \tilde{x}  =  \frac{x}{1 + \xi^2}
\end{eqnarray}
where $E_e$ is the electron beam energy and $E_0$ the energy of
the laser photon.
As a result, the energy distribution for the backscattered photons 
is shifted towards lower energies.

\subsection{Angular correlations}

Electron beams collide with focused laser beams 
at the distance $b\sim 2$mm
from the interaction point.
The angular spread of scattered photons is very
small (due to very high Lorenz boost), but becomes
important because of the very small beam spot size.
Due to the larger scattering angle, the effective vertical size of
the photon beam increases at low photon energies. 
As a result, interactions involving low energy photons 
are suppressed.
The effect of angular correlations
can be described by the following modification 
of the energy spectrum \cite{Ilya}:
\begin{eqnarray}
\frac{1}{N} \frac{dN}{dy} = f(y,\tilde{x})  & = & 
  f_C(y,\tilde{x}) \cdot {\cal N } \exp \left(-\frac{\rho^2}{8} \; 
\left(\frac{\tilde{x}}{y} -\tilde{x} -1 \right) \right) \; ,
\label{eq:compang}
\end{eqnarray}
where the parameter $\rho$ relates the vertical beam size $\sigma_y$ and 
the distance $b$ between the conversion and interaction points,
$\rho \approx (m_e b)/(E_e \sigma_y)$,   ${\cal N}$ is normalization factor
and $f_C(y,\tilde{x})$ is the Compton spectrum, as described 
by eq. (\ref{eq:compt}) in appendix \ref{sec:compton}.

\subsection{Electron rescattering}

With high density of the laser beam, 
one can "convert" most of the electrons into the
high energy photons. 
However, after the first scattering electrons still have large
energies.
Scattering of  laser photons on these secondary electrons
results in an additional contribution to the photon-energy
spectrum.
Due to the lower electron energy, 
secondary scattering takes place at lower $x$ value, $x' = y' \tilde{x}$,
where $y'$  is the energy fraction of the secondary electrons.
Energy distribution for photons scattered off secondary electrons
can be calculated by integrating over $y'$:
\begin{eqnarray}
\frac{1}{N'} \frac{dN'}{dy} = f'(y,\tilde{x}) & = &  {\cal N}'
\int\limits_{0}^{1} dy^\prime \;\;\; w_C(y^\prime) 
  \;  f_C(1-y^\prime,\tilde{x}) \cdot 
      f_C(\frac{y}{y^\prime},y^\prime \tilde{x})
\label{eq:sece}
\end{eqnarray}
where ${\cal N}'$ is the normalization factor
and $w_C(y')$ is the weighting function which takes into account
the dependence of the total Compton scattering cross section on 
the electron energy \cite{GKST83,compton}. In the high energy limit 
one gets
\begin{eqnarray}
w_C(y^\prime) & \approx & \frac{\tilde{x}}{x'} \cdot
\frac{\log (x'+1)}{\log (\tilde{x}+1)} \; .
\label{eq:wc}
\end{eqnarray}
Photons from scattering on secondary electrons 
have much ``softer'' energy spectrum, as compared to the
Compton scattering on primary electrons.

\subsection{Scattering of two laser photons}

For high density of laser beam it is also possible that electron
scatters on two photons instead of one. 
The detailed calculation
of the energy distribution for photons produced in such scattering 
is presented in \cite{Galynskii}.
We have found that the distribution obtained from the simulation
 \cite{TEL01} can be well approximated
by a simple formula for the scattering on one photon with double energy
(i.e. with $\tilde{x} \rightarrow 2\tilde{x}$) corrected by
an additional factor which suppresses the high energy peak:
\begin{eqnarray}
\frac{1}{N_2} \frac{dN_2}{dy} = f_2(y,\tilde{x})  & = & f(y,2 \tilde{x}) \cdot 
{\cal N}_2 \; \left(\frac{2 \tilde{x}}{y} - 2 \tilde{x} -1 \right)^\delta \; ,
\label{eq:comptwo}
\end{eqnarray}
where $\delta$ is the parameter describing suppression of the high 
energy part of the spectrum and ${\cal N}_2$ is the normalization constant.


\section{Parametrization}
\label{sec:param}

\subsection{Main assumptions}

The main aim of the presented study was to parametrize the high energy
part of the luminosity spectra of the Photon Collider in a simple
analytical form.
It was assumed that the high energy part of the \gaga\ luminosity spectrum
can be described as a simple product:
\begin{eqnarray}
\frac{1}{N} \frac{d^2 N}{dy_1 dy_2} & = & 
           f_{tot}(y_1,\tilde{x}) \; f_{tot}(y_2,\tilde{x}) 
\label{eq:f2d}
\end{eqnarray}
where $f_{tot}(y,\tilde{x})$ is the energy spectrum for the photon.
The spectrum can be parametrized as a sum of
three components described in the previous section:
\begin{eqnarray}
f_{tot}(y,\tilde{x}) & = & n\; f(y,\tilde{x}) 
                  \; + \; n' \; f'(y,\tilde{x}) 
                  \; + \; n_2 \; f_2(y,\tilde{x}) 
\label{eq:compaz}
\end{eqnarray}
where $n$, $n'$ and $n_2$ are parameters describing contributions
of different processes to the spectrum.
All together the model has 10 free parameters which can be
adjusted to describe the results of simulation by Telnov \cite{TEL01}. 
Only 4 of these parameters describe the shape of the contributing components:
\begin{itemize}
  \item two parameters,  $\xi_0$ and $\xi_1$, 
 describing $\xi^2$ dependence on the electron beam energy: 
\begin{eqnarray}
            \xi^2 & = & \xi_0 + E_e \cdot \xi_1
\end{eqnarray}
  \item parameter $\rho^2$ describing the angular correlations 
  \item parameter $\delta$ added in description of two-photon scattering
\end{itemize}
Remaining  6 parameters, $a_1 \ldots a_6$, 
are needed to describe the normalization of the contributing processes:
\begin{itemize}
  \item the dependence of the normalization of the 
  Compton scattering contribution  ($n$)
on the electron beam energy is, in the considered energy range, 
 approximately given by:
\begin{eqnarray} 
    n\: {\cal N} = a_1 \cdot E_e \; + \; a_2
\end{eqnarray}
  \item normalization of the electron rescattering ($n'$) and
of the two photon scattering ($n_2$) 
are related to the Compton scattering contribution by the formula: 
\begin{eqnarray}
n' \: {\cal N}'  & =  & 
      n \: {\cal N} \cdot \left( a_3 \cdot E_e \; + \; a_4 \right) \\
n_2 \: {\cal N}_2  & = & n \: {\cal N} \cdot a_5
  \left( 1 -\exp (-a_6\: \xi^2 ) \right)
\end{eqnarray}
\end{itemize}
Normalization factors  ${\cal N}$, ${\cal N}'$ and ${\cal N}_2$ 
are included in the parametrization to simplify
numerical calculations of the spectra.

Whereas distributions $f$, $f'$ and $f_2$ are normalized
to unity, normalization of $f_{tot}$ is not fixed.
It is evaluated from the requirement that 
the high energy part of the luminosity spectrum 
is given by the formula  (\ref{eq:f2d}).

\subsection{Fit results}
\label{sec:fit}

The formula  (\ref{eq:compaz}) was compared with
the photon energy spectra obtained from simulation
by V.Telnov \cite{TEL01}.
To minimize effects of energy correlations 
a cut on the energy of the second photon was imposed.
For electron beam energy of 100, 250 and 400 GeV
the cut was 40, 150 and 260 GeV, respectively.
Parameters of the model were fitted
to the photon spectra, for $y>0.1$, simultaneously at all energies.

\begin{figure}[tbp]
  \begin{center}
  \epsfig{figure=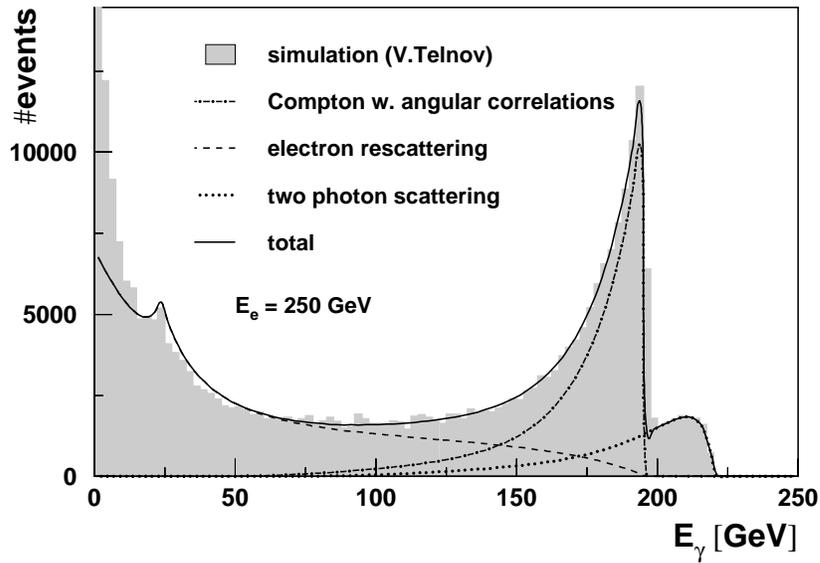,height=\figheight,clip=}
  \end{center}
  \caption{Comparison of the  photon energy distribution 
        obtained from full simulation of luminosity spectrum 
        by Telnov \cite{TEL01},
        with the fitted contributions of different processes
        considered in the described model, as indicated in the plot.
           }
  \label{fig:fitcontr}
\end{figure}

\begin{figure}[tbp]
  \begin{center}
  \epsfig{figure=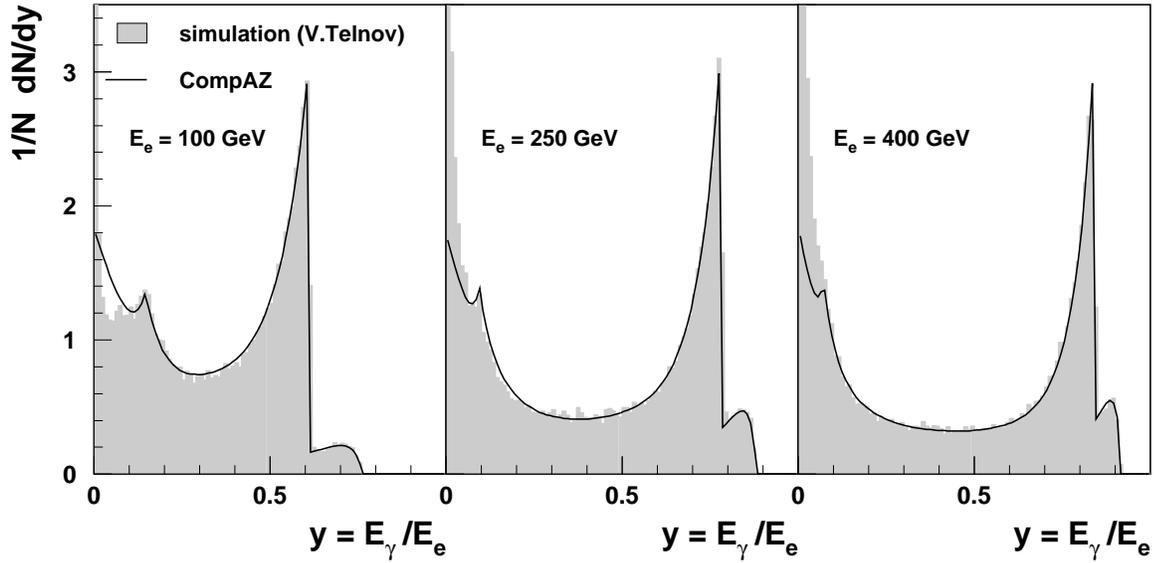,width=\figwidth,clip=}
  \end{center}
  \caption{Comparison of the  photon energy distribution 
        from the fitted parametrization with the distribution
        obtained from full simulation of luminosity spectra \cite{TEL01},
        for three electron beam energies, as indicated in the plot.
        Imposed cut on the energy of the second photon is 
        40, 150 and 260 GeV respectively.
           }
  \label{fig:fitene}
\end{figure}

Result of the fit to the photon energy distribution
at $E_e$=250 GeV is shown in Fig.~\ref{fig:fitcontr}.
Fitted contributions of different processes
are also indicated.
The model describes the spectra very well down
to $E_\gamma \sim 0.1 \; E_e$. 
Three processes considered in the model contribute to
different parts of the spectra.
By summing these contributions
most of details of the distribution can be well reproduced.
Very good description of the photon energy distribution 
is obtained for all considered energies, as shown 
in  Fig.~\ref{fig:fitene}.

\begin{figure}[tbp]
  \begin{center}
  \epsfig{figure=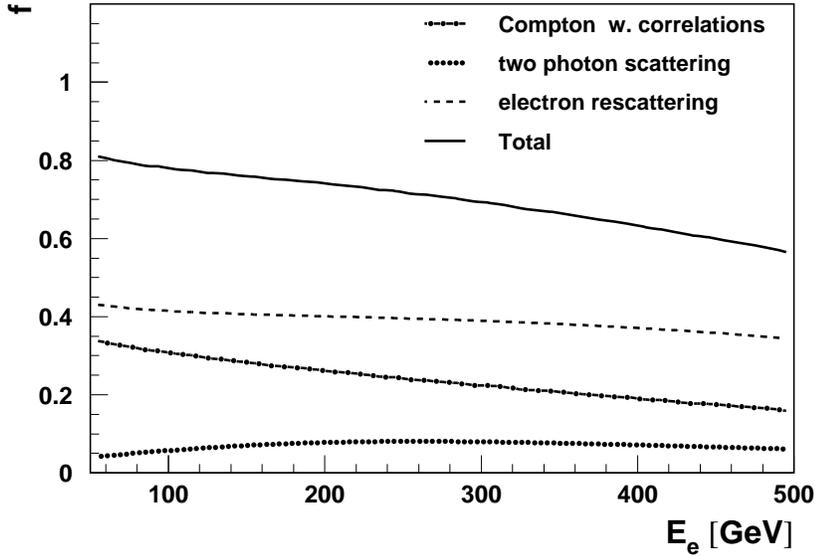,height=\figheight,clip=}
  \end{center}
  \caption{Normalization of the CompAZ parametrization of
        the photon energy distribution, relative to the
        distribution obtained from full simulation of 
        luminosity spectrum \cite{TEL01}, 
        as a function of the electron beam energy.
        Also shown are normalizations of separate processes
        considered in the model.
           }
  \label{fig:norm}
\end{figure}

Normalization of the fitted parametrization, as well
as of the contributions of different processes,
are shown in Fig.~\ref{fig:norm} as a function of
the electron beam energy $E_e$.
Normalization of the parametrization changes from about 0.8
at 50 GeV to about 0.55 at 500 GeV.
This means that the two photon spectrum obtained from the product 
of the two distributions, as given by eq.~(\ref{eq:f2d}),
describes between 65\% and 30\% of events expected from 
the spectra simulation \cite{TEL01}.
35\% to 70\% of the total \gaga\ luminosity expected from
simulation is due to events with one or two low energy photons 
not described by our parametrization.

\begin{figure}[tbp]
  \begin{center}
  \epsfig{figure=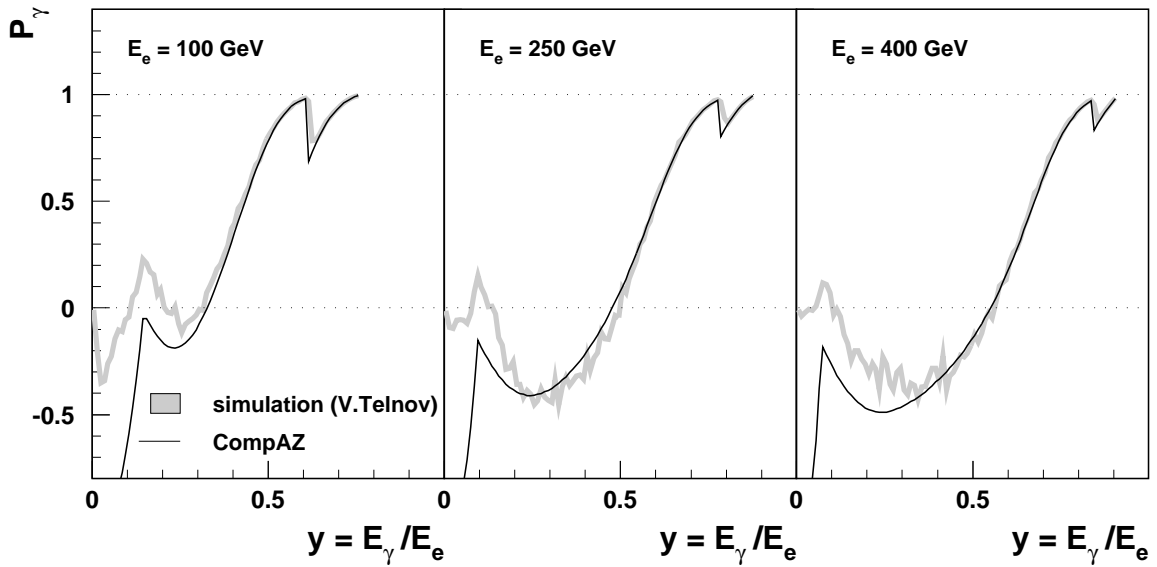,width=\figwidth,clip=}
  \end{center}
  \caption{Comparison of the  photon polarization resulting
        from the fitted parametrization with the distribution
        obtained from full simulation of luminosity spectra \cite{TEL01},
        for three electron beam energies, as indicated in the plot.
        Imposed cut on the energy of the second photon is 
        40, 150 and 260 GeV respectively.
           }
  \label{fig:fitpol}
\end{figure}

In Fig.~\ref{fig:fitpol} the comparison of the 
average photon polarization resulting from the fitted 
parametrization with the distribution obtained from 
the simulation of luminosity spectra is shown.
To describe the photon polarization
two additional assumptions were made in the model: 
\begin{itemize}
\item scattering involving two photons results in very 
high photon polarization. It is taken from the Compton formula (\ref{eq:polar})
(with $\tilde{x} \rightarrow 2 \tilde{x}$) for scattered photon energies 
above the threshold for one photon scattering, and fixed at the threshold
value for lower energies;
\item 
electrons undergoing secondary scattering are 
unpolarized.
\end{itemize}
Both assumptions have no strong physical motivation,\footnote{
As it was pointed out by V.Telnov, significant polarization 
is expected for secondary electrons \cite{Galynskii}.}
however they were found to give the best description
at the simulation level.
It has to be stressed that the model was not fitted to the 
photon polarization distribution and no additional parameters were
introduced to describe it. 
Very good agreement between the parametrization 
and the average photon polarization obtained from the simulation,
is observed for $y > 0.3$.

\subsection{CompAZ}
\label{sec:compaz}

The routine implementing the described spectra parametrization
is called {\bf CompAZ}. It can be used to calculate the photon
energy spectrum for different electron beam energies and
the average photon polarization for a given photon
energy.
Separate contributions from three considered processes 
(\ref{eq:compang},\ref{eq:sece},\ref{eq:comptwo})
can also be calculated.
Additional routines were prepared 
for convenient event generation from the parametrized spectrum.
All routines can be downloaded from web \cite{compaz}.

\begin{figure}[tbp]
  \begin{center}
  \epsfig{figure=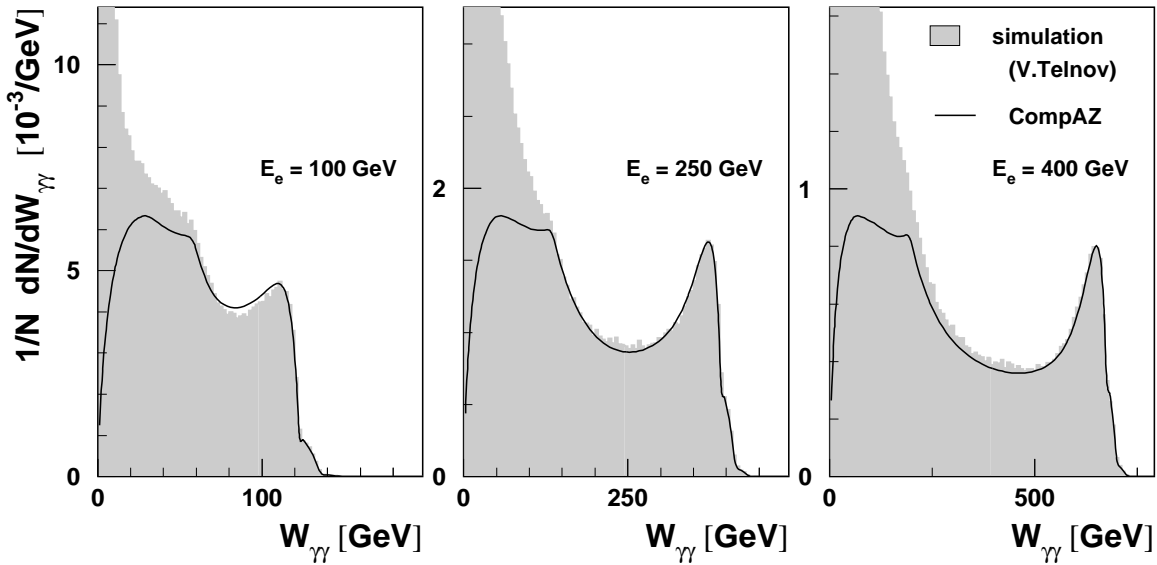,width=\figwidth,clip=}
  \end{center}
  \caption{Comparison of the center-of-mass energy distribution 
         obtained with the CompAZ parametrization with the distribution
        obtained from full simulation of luminosity spectra \cite{TEL01},
        for three electron beam energies, as indicated in the plot.
           }
  \label{fig:fitw}
\end{figure}

\begin{figure}[tbp]
  \begin{center}
  \epsfig{figure=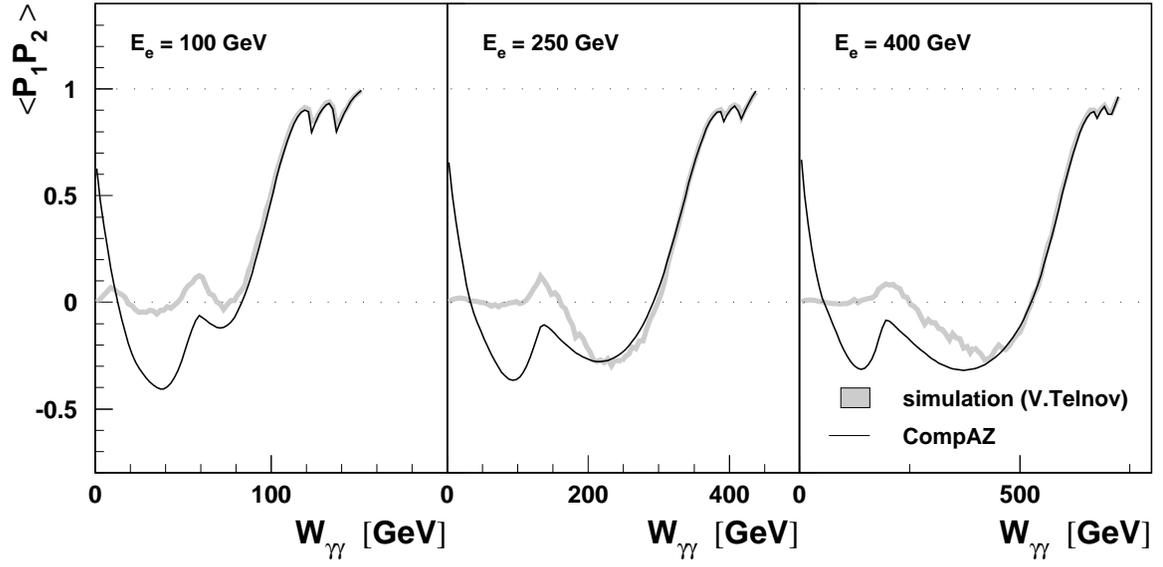,width=\figwidth,clip=}
  \end{center}
  \caption{Comparison of the average product of photon polarizations
        from the CompAZ parametrization with the distribution
        obtained from full simulation of luminosity spectra \cite{TEL01},
        for three electron beam energies, as indicated in the plot.
           }
  \label{fig:fitp2}
\end{figure}

\begin{figure}[tbp]
  \begin{center}
  \epsfig{figure=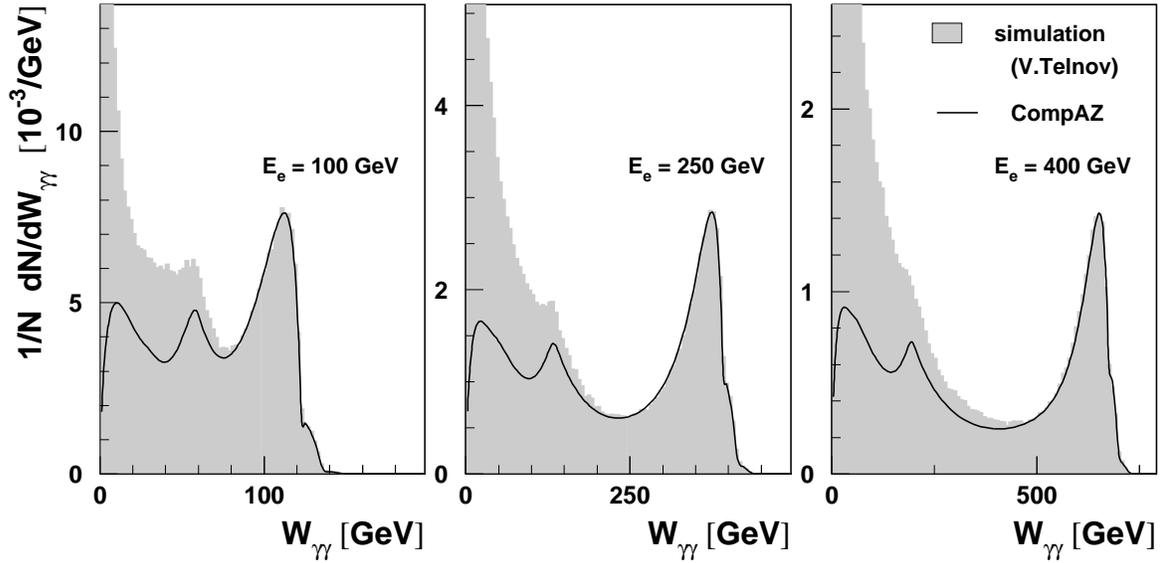,width=\figwidth,clip=}
  \end{center}
  \caption{Center-of-mass energy distribution 
        for two colliding photons with  $J_z=0$. 
        Results obtained with the CompAZ parametrization
        are compared with the distribution
        obtained from full simulation of luminosity spectra \cite{TEL01},
        for three electron beam energies, as indicated in the plot.
           }
  \label{fig:fitw0}
\end{figure}

In Fig.~\ref{fig:fitw} the \gaga\ center-of-mass energy distribution 
obtained from  CompAZ is compared with the distribution
obtained from the simulation of luminosity spectra \cite{TEL01}, for 
different electron beam energies.
No cuts on photon energies were imposed.
Proper description of the spectra is obtained for
$W_{\gamma \gamma}   >  \sim 0.3\; W_{max}$, where 
$W_{max}  = 2\; E_{max}$ is the maximum center-of-mass energy
available for two photons produced in the 
Compton scattering.
Also the average product of the photon polarizations,
related to the ratio of \gaga\ collisions with the total
angular momentum $J_z=0$ and $|J_z|=2$,
is properly described for large $W_{\gamma \gamma}$ (see Fig.~\ref{fig:fitp2}).
Center-of-mass energy distribution 
for two colliding photons with  $J_z=0$ is shown in Fig.~\ref{fig:fitw0}.
The parametrization describes very well the high energy part
of the spectra,  most relevant for many physics studies.

\begin{figure}[tbp]
  \begin{center}
 \epsfig{figure=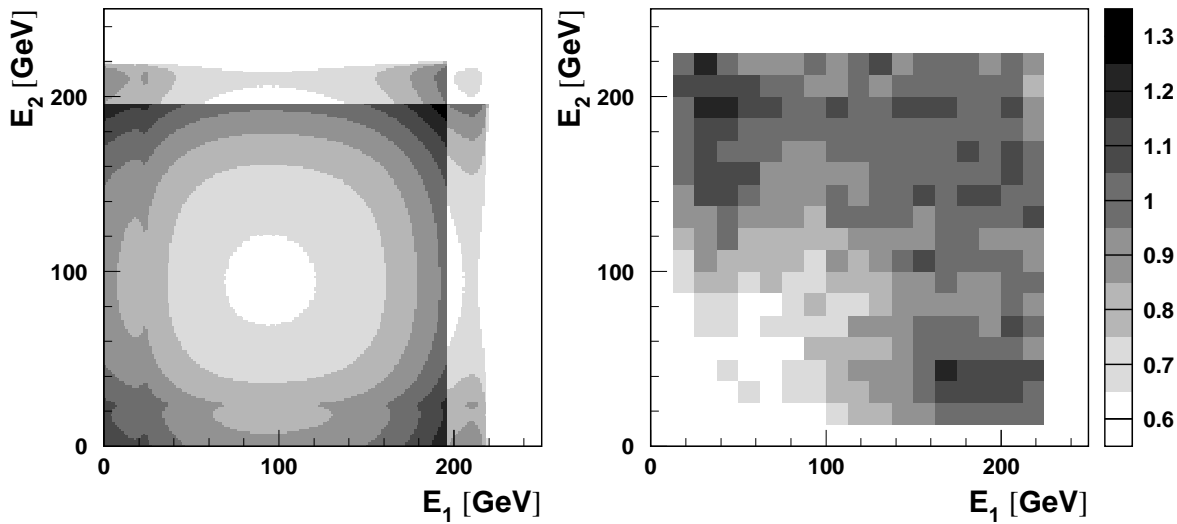,width=\figwidth,clip=}
  \end{center}
  \caption{Two-dimensional energy distribution for two colliding photons, 
           obtained from the CompAZ parametrization (left plot) and
           the ratio of this distribution to 
           the one obtained from full simulation of luminosity 
           spectrum \cite{TEL01} (right plot).
           }
  \label{fig:compaz2}
\end{figure}

In the calculation of the \gaga\ luminosity spectrum as 
a product of two energy distributions (\ref{eq:f2d})
possible energy correlations between two beams are neglected.
In Fig.~\ref{fig:compaz2} the
two-dimensional energy distribution
obtained from CompAZ,
and the ratio of this distribution to 
the two-photon spectrum obtained from the simulation by V.Telnov \cite{TEL01}
are shown.
In the high energy part of the spectrum,
when both photons have high energies, $y>0.5 \; y_{max}$, 
this ratio is  close to 1.
This shows that CompAZ properly describes this part of the spectrum and
no additional corrections for beam energy correlations are needed.
Only for $W_{\gamma \gamma}   <  \sim 0.3\; W_{max}$ energy
correlations become important. 
CompAZ overestimates the number of collisions 
with one high energy ($y>0.5 \; y_{max}$) and 
one low energy ($y<0.3 \; y_{max}$) photon 
(the ratio greater than 1),
and underestimates the number of collisions involving two low
energy photons (the ratio smaller than 1).
  
\subsection{Applications}
\label{sec:appl}

\begin{figure}[tbp]
  \begin{center}
  \epsfig{figure=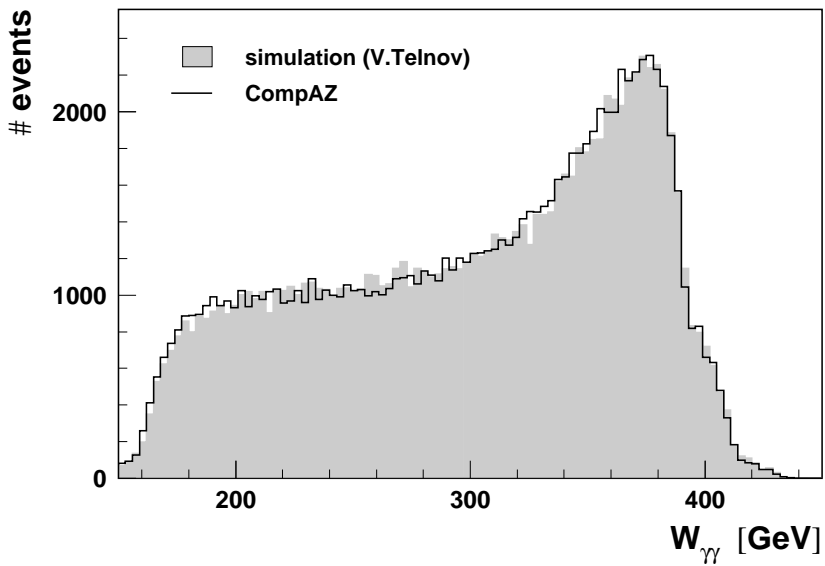,height=\figheight,clip=}
  \end{center}
  \vspace{-0.5cm}
  \caption{Distribution of the $\gamma \gamma$ center-of-mass energy 
         for $\gamma \gamma \rightarrow W^+ W^-$ events 
        generated with PYTHIA, for electron beam energy of 250 GeV.
        Sample of events generated using
        the CompAZ parametrization is compared with the sample
        generated with the luminosity spectrum from full simulation.
        Generated events were reweighted for photon polarizations.
           }
  \label{fig:wgen}
\end{figure}

As already mentioned in the previous section, dedicated routines are available
for fast simulation of \gaga\ scattering events with CompAZ. 
They have been recently used in the simulation of $W^+ W^-$ 
and $ZZ$ pair-production
at the Photon Collider, for different electron beam energies \cite{ichep_ww}.
Distribution of the $\gamma \gamma$ center-of-mass energy,
$W_{\gamma \gamma}$ for 
$\gamma \gamma \rightarrow W^+ W^-$ events generated 
with PYTHIA, for electron beam energy of 250 GeV,
is shown in Fig.~\ref{fig:wgen}.
Generated events were reweighted for photon polarizations.
Sample of events generated using CompAZ is compared with the sample
generated with the luminosity spectrum from simulation \cite{TEL95,TEL01}.
Very good agreement is observed.
The advantage of CompAZ is that it can be easily used for any
beam energy\footnote{Parametrization can be used for $50<E_e < 500$ GeV.},
giving reasonable description of the energy and polarization.

\begin{figure}[tbp]
  \begin{center}
  \epsfig{figure=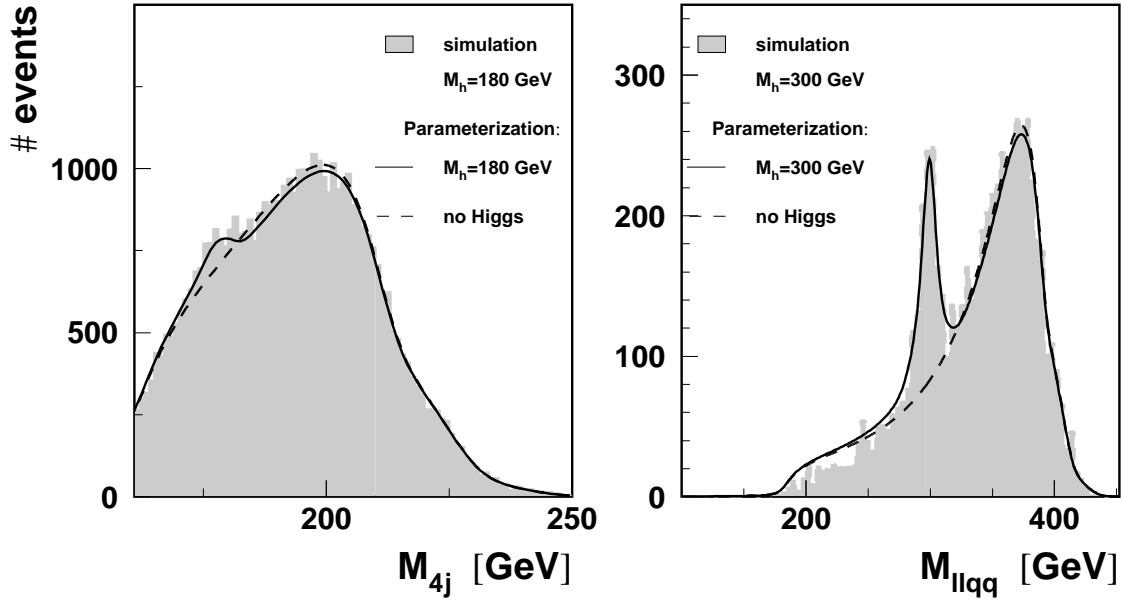,width=\figwidth,clip=}
  \end{center}
  \caption{Distribution of the reconstructed invariant mass
        for $\gamma \gamma \rightarrow W^+ W^-$ events with a 
        SM Higgs-boson mass of 180 GeV and an 
        electron-beam energy of 152.5 GeV (left plot) 
        and
        for $\gamma \gamma \rightarrow Z  Z $ events, with a 
        SM Higgs-boson mass of 300 GeV and an 
        electron-beam energy of 250 GeV (right plot). 
        Results from the simulation  based  on PYTHIA 
        and on the SIMDET detector simulation (histogram) are 
        compared with the distribution
        obtained  by the numerical convolution of the cross-section 
        formula with the CompAZ photon energy spectra
        and parametrization of the detector resolution (solid line).
        The distribution expected without the Higgs contribution is
        also shown (dashed line).
          }
  \label{fig:whgen}
\end{figure}

CompAZ parametrization can also be used for calculating expected
event distributions without the time consuming event generation.
Numerical integration is few orders of magnitude faster than the full
simulation and can be used to extrapolate results of full simulation
to other beam energies.
Results from the recent study of the heavy Higgs boson
production at the Photon Collider \cite{ichep_ww},
for Higgs mass of 180 GeV ($h \rightarrow W^+ W^- $)
and 300~GeV ($h \rightarrow ZZ$) are shown in Fig.~\ref{fig:whgen}.
Expected invariant mass distributions obtained
from the full simulation, based on PYTHIA \cite{PYTHIA} and
fast detector simulation program SIMDET \cite{SIMDET}, are compared with 
results obtained  by the numerical convolution of the cross section 
formula for vector boson production in \gaga\ scattering
with the CompAZ photon energy spectrum
and parametrization of the detector resolution.
The agreement between both approaches is very good.
With numerical integration based on CompAZ,
it was possible to calculate the detector level effects 
expected from the interference between direct $W^+ W^-$ production and 
$h \rightarrow W^+ W^-$ decay.
To estimate the effect with full event and detector simulation
very large statistic of events would be required.


\section{Summary}
\label{sec:summary}

Luminosity spectrum obtained from the detailed beam simulation 
is the best tool for accurate simulation of $\gamma \gamma$ 
interactions.
Recently the new version of CIRCE code became available \cite{CIRCE1,CIRCE2},
which contains the Photon Collider luminosity spectra based on 
simulation by V.Telnov \cite{TEL95,TEL01}.
The CIRCE program \cite{CIRCE2} gives 
detailed description of the luminosity spectra
in the whole energy range taking properly into account all
non-factorizing contributions (energy and polarization correlations).
The package includes routines for convenient event generation.
However, only three selected electron beam energies have been 
considered so far ($E_e$= 100, 250 and 400 GeV).
Moreover, parameters of the Photon Collider assumed in the simulation
are only known with accuracy up to 10--20\%, 
as many details of the project are still not fixed.
Therefore, other models resulting in the similar (or better) accuracy
are also applicable for detailed studies.

We propose the model describing the photon energy spectra 
of the Photon Collider at TESLA in a simple analytical form,
based on the formula for the Compton scattering.
Parameters of the model are obtained from the comparison
with the full beam simulation by V.Telnov, 
which includes nonlinear corrections and
contributions of higher order processes. 
Photon energy distribution and 
polarization, in the high energy part of the spectra, are well reproduced
in a wide range of electron beam energies.
Model can be used for Monte Carlo simulation of gamma-gamma events.
Parametrization is also very useful for
numerical cross-section calculations.

\section*{Acknowledgments}

Special thanks are due to V.Telnov for providing the
results of the Photon Collider luminosity spectra simulation
and to M.Krawczyk for many productive discussions 
and valuable comments.
I would also like to thank V.Telnov, I.Ginzburg and T.Ohl
for comments and critical remarks to this paper.


\clearpage

\appendix

\section{Energy spectrum for Compton scattering}
\label{sec:compton}

The energy spectrum of the photons resulting from the Compton backscattering
of laser light off the high energy electron beam 
depends on the electron beam and laser polarizations, $P_e$ and $P_L$,
and on the dimension--less parameter $x$:
\begin{eqnarray}
x &=&  \frac{4 E_0 E_e}{m_e^2},
\label{eq:x}
\end{eqnarray}
where $E_e$ is the electron beam energy, $E_0$ the energy of
the laser photon and $m_e$ is the electron mass.
The maximum energy of the scattered 
photon is 
\begin{eqnarray}
E_{max} &=& \frac{x}{x+1}\; E_e.
\end{eqnarray}
and the energy spectrum is given by \cite{GKST84}
\begin{eqnarray}
\frac{1}{N} \frac{dN}{dy} & = & f_C (y,x) = \nonumber \\  & = & \!
{\cal N}_C \left[\frac{1}{1-y}-y+(2r-1)^2 - \! P_e P_L \;
xr(2r-1)(2-y)\right] , 
\label{eq:compt}
\end{eqnarray}
where $r =y/(x(1-y))$, $y$ is the fraction of the electron energy 
transfered to the photon
\begin{eqnarray}
0 \; \le \; y  =  \frac{E_{\gamma}}{E_e}  &  \le & \frac{x}{x+1} \; ,
\end{eqnarray}
and ${\cal N}_C$ is the normalization factor given by
\begin{eqnarray}
\frac{1}{{\cal N}_C } & = &
   \frac{1}{2} + \frac{8}{x} - \frac{1}{2(x+1)^2} +
   \left(1 - \frac{4}{x} - \frac{8}{x^2}\right)\log(1+x) \nonumber \\
& - &
   P_e P_L \left[ 2  +  \frac{x^2}{2(x+1)^2} - 
           \left(1+\frac{2}{x}\right)\log(1+x) \right] \; . 
\label{eq:cnorm}
\end{eqnarray}

The degree of polarization of the photons scattered with energy
fraction $y$ is given by \cite{GKST84}  
\begin{eqnarray}
P_\gamma \! & = & \! \frac{\cal N_C}{f_C (y,x)} 
 \left\{  x r P_e  \left[1+(1-y)(2r-1)^2\right]
- (2r-1) P_L \left[\frac{1}{1-y}+1-y\right] \right\}.
\nonumber \\ & &
\label{eq:polar}
\end{eqnarray}

\begin{figure}[tbp]
  \begin{center}
  \epsfig{figure=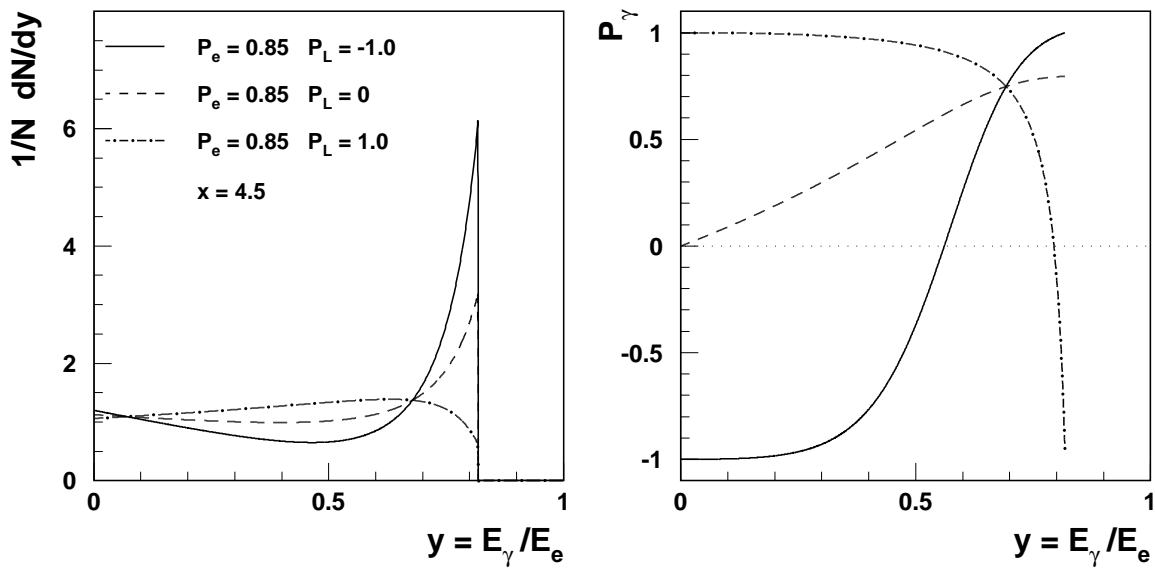,width=\figwidth,clip=}
  \end{center}
  \caption{Energy distribution (left plot) and polarization (right plot) 
           for photons from Compton back-scattering,
           for different laser beam polarizations $P_L$, 
           as indicated in the plot.
           $x=4.5$ corresponds to laser wave length of 1.06 $\mu m$ and
           primary electron beam energy of 250 GeV.
           Electron beam polarization  $P_e$=85\%.}
  \label{fig:compton}
\end{figure}

The energy spectrum and polarization of the scattered photons, for $x=4.5$ 
and 85\% polarization of the electron beam 
(proposed parameters of the TESLA Photon Collider for
electron beam energy $E_e$=250 GeV), 
and various helicities of laser beam are shown in Fig.~\ref{fig:compton}.

\clearpage
 
%
%


\def\baselinestretch{0.7}


\end{document}